\RequirePackage{cmap}%
\documentclass{scrartcl}
\pdfoutput=1

\usepackage[utf8]{inputenc}
\usepackage[T1]{fontenc}
\usepackage{ifpdf}
\usepackage{cite}
\usepackage{caption}

\usepackage[charter]{mathdesign}
\usepackage{berasans,beramono}
\usepackage[final,tracking=true,kerning=true,spacing=true]{microtype}
\microtypecontext{spacing=nonfrench}

\usepackage{amsmath}

\usepackage{algorithm}
\usepackage[noend]{algpseudocode}

\usepackage{etoolbox}
\newtoggle{haspgfplotsstatistics}
\newtoggle{externalizeplots}

\togglefalse{haspgfplotsstatistics}
\togglefalse{externalizeplots}

\iftoggle{haspgfplotsstatistics}{
  \usepackage{pgfplots}
  \pgfplotsset{compat=1.9}
  \usetikzlibrary{plotmarks}
  \usepgfplotslibrary{statistics}
  \iftoggle{externalizeplots}{
    \usepgfplotslibrary{external}
    \tikzexternalize
  }{}
}{
  \usepackage{graphicx}
}

\usepackage{pgfplotstable}
\usepackage{booktabs}

\pgfkeys{/pgf/number format/1000 sep={\,}}
\iftoggle{haspgfplotsstatistics}{
\pgfplotsset{
  boxplot/draw direction=y,
  cycle list name=black white,
  every axis plot/.style={
    colormap name=blackwhite
  },
  every semilogy axis/.style={
    cycle list={{mark=-}},
    log ticks with fixed point,
    scaled ticks=false,
  },
}}{}
\pgfplotstableset{
  col sep=comma,
  every head row/.style={
    before row=\toprule,
    after row=\midrule,
  },
  every last row/.style={
    after row=\bottomrule,
  },
}

\newcommand{\mlcell}[2][c]{\begin{tabular}[t]{@{}#1@{}}#2\end{tabular}}
\newcommand*{\bytesToMiB}[1]{
  \pgfkeyssetvalue{/pgfplots/table/@cell content}{%
    \pgfkeysalso{/pgf/number format/fixed}%
    \pgfkeysalso{/pgf/number format/precision=1}%
    \pgfmathfloatparsenumber{1048576}%
    \let\mydivisor\pgfmathresult%
    \pgfmathfloatparsenumber{#1}%
    \pgfmathfloatdivide{\pgfmathresult}{\mydivisor}%
    \pgfmathprintnumber{\pgfmathresult}\,MiB%
  }%
}%

\algrenewcommand\algorithmicrequire{\textbf{Input:}}
\algrenewcommand\algorithmicensure{\textbf{Output:}}
\algrenewcommand\algorithmicforall{\textbf{for each}}
\newcommand*{\ForEach}{\ForAll}
\newcommand*{\mleft}{\mathopen{}\mathclose\bgroup\left}
\newcommand*{\mright}{\aftergroup\egroup\right}
\newcommand*{\targetstop}{p_\mathrm{tgt}}
\newcommand*{\sourcestop}{p_\mathrm{src}}
\newcommand*{\chTime}[1]{\Delta\tau_{\mathrm{ch}}\mleft(#1\mright)}
\newcommand*{\fpTime}[2]{\Delta\tau_{\mathrm{fp}}\mleft(#1,#2\mright)}
\newcommand*{\arrTime}[2]{\tau_{\mathrm{arr}}\mleft(#1,#2\mright)}
\newcommand*{\depTime}[2]{\tau_{\mathrm{dep}}\mleft(#1,#2\mright)}
\newcommand*{\seq}[1]{\mleft<#1\mright>}
\newcommand*{\stopsOn}[1]{\vec{p}\mleft(#1\mright)}
\newcommand*{\nthStopOn}[2]{p^{#2}_{#1}}
\newcommand*{\lineOf}[1]{L_{#1}}
\newcommand*{\linesAt}[1]{\boldsymbol{L}\mleft(#1\mright)}
\newcommand*{\tripSegment}[3]{p^{#2}_{#1} \rightarrow p^{#3}_{#1}}
\newcommand*{\transfer}[4]{p^{#2}_{#1} \rightarrow p^{#4}_{#3}}
\newcommand*{\reached}[2][]{R_{#1}\mleft(#2\mright)}

\ifpdf
\fi
\begin{document}
\title{Trip-Based Public Transit Routing}
\author{Sascha Witt \\ \texttt{sascha.witt@kit.edu} \\ \\
  Karlsruhe Institute of Technology (KIT)\\Karlsruhe, Germany}
\date{}
\maketitle
\begin{abstract}
We study the problem of computing all Pareto-optimal journeys in a public transit network regarding the two criteria of arrival time and number of transfers taken.
We take a novel approach, focusing on trips and transfers between them, allowing fine-grained modeling.
Our experiments on the metropolitan network of London show that the algorithm computes full $24$-hour profiles in $70$\,ms after a preprocessing phase of $30$\,s, allowing fast queries in dynamic scenarios.
\end{abstract}
\section{Introduction}\label{sec:introduction}

Recent years have seen great advances in route planning on continent-sized road networks~\cite{Bast2014}.
Unfortunately, adapting these algorithms to public transit networks is harder than expected~\cite{Berger2009}.
On road networks, one is usually interested in the shortest path between two points, according to some criterion.
On public transit networks, several variants of point-to-point queries exist.
The simplest is the \emph{earliest arrival query}, which takes a departure time as an additional input and returns a journey that arrives as early as possible.
A natural extension is the \emph{multi-criteria} problem of minimizing both arrival time and the number of transfers, resulting in a set of journeys.
A \emph{profile query} determines all optimal journeys departing during a given period of time.

In the past, these problems have been solved by modeling the timetable information as a graph and running Dijkstra's algorithm or variants thereof on that graph.
Traditional graph models include the time-expanded and the time-dependent model~\cite{Muller-Hannemann2007}.
More recently, algorithms such as RAPTOR~\cite{Delling2012} and Connection~Scan~\cite{Dibbelt2013} have eschewed the use of graphs (and priority queues) in favor of working directly on the timetable.

In this work, we present a new algorithm that uses trips (vehicles) and the transfers between them as its fundamental building blocks.
Unlike existing algorithms, it does not assign labels to stops.
Instead, trips are labeled with the stops at which they are boarded.
Then, a precomputed list of transfers to other trips is scanned and newly reached trips are labeled.
When a trip reaches the destination, a journey is added to the result set.
The algorithm terminates when all optimal journeys have been found.

A motivating observation behind this is the fact that labeling stops with arrival (or departure) times is not sufficient once minimum change times are introduced.
Some additional information is required to track which trips can be reached.
For example, the realistic time-expanded model of Pyrga et~al.~\cite{Pyrga2008} introduces additional nodes to deal with minimum change times, while Connection~Scan~\cite{Dibbelt2013} uses additional labels for trips.
In contrast, once we know passengers boarded a trip at a certain stop, their further options are fully defined: Either they transfer to another trip using one of the precomputed transfers, or their current trip reaches the destination, in which case we can look up the arrival time in the timetable.
In either case, there is no need to explicitly track arrival times at intermediary stops.

The core of the algorithm is similar to a breadth-first search, where levels correspond to the number of transfers taken so far.
As a result, it is inherently multi-criterial, similar to RAPTOR~\cite{Delling2012}.
Although a graph-like structure is used, there is no need for a priority queue.
A preprocessing step is required to compute transfers, but can be parallelized trivially and only takes minutes, even on large networks (Section~\ref{sec:experiments}).
By omitting unnecessary transfers, space usage and query times can be improved at the cost of increased preprocessing time.

Section~\ref{sec:preliminaries} introduces necessary notations and definitions, before Section~\ref{sec:algorithm} describes the algorithm and its variants.
Section~\ref{sec:experiments} presents the experimental evaluation.
Finally, Section~\ref{sec:conclusions} concludes the paper.

\section{Preliminaries}\label{sec:preliminaries}
\subsection{Notation}

We consider public transit networks defined by an aperiodic \emph{timetable}, consisting of a set of stops, a set of footpaths and a set of trips.
A \emph{stop} $p$ represents a physical location where passengers can enter or exit a vehicle, such as a train station or a bus stop.
Changing vehicles at a stop $p$ may require a certain amount of time $\chTime{p}$ (for example, in order to change platforms).%
\footnote{More fine-grained models, such as different change times for specific platforms, can be used without affecting query times, since minimum change times are only relevant during preprocessing (Section~\ref{sec:preprocessing}).}
\emph{Footpaths} allow travelers to walk between two stops.
We denote the time required to walk from stop $p_1$ to $p_2$ by $\fpTime{p_1}{p_2}$ and define $\fpTime{p}{p} = \chTime{p}$ to simplify some algorithms.
A \emph{trip} $t$ corresponds to a vehicle traveling along a sequence of stops $\stopsOn{t} = \seq{\nthStopOn{t}{0}, \nthStopOn{t}{1}, \dotsc}$.
Note that stops may occur multiple times in a sequence.
For each stop $\nthStopOn{t}{i}$, the timetable contains the arrival time $\arrTime{t}{i}$ and the departure time $\depTime{t}{i}$ of the trip at this stop.
Additionally, we group trips with identical stop sequences into \emph{lines}%
\footnote{We chose \emph{line} over \emph{route} to avoid confusion with \emph{routing} and the usage of \emph{route} in the context of road networks.}
such that all trips $t$ and $u$ that share a line can be totally ordered by
\begin{equation}\label{eq:trip_order}
  t \preceq u \iff \forall i \in \mleft[0, \mleft|\stopsOn{t}\mright|\mright):
  \arrTime{t}{\nthStopOn{t}{i}} \leq \arrTime{u}{\nthStopOn{u}{i}}
\end{equation}
and define
\begin{equation}
  t \prec u \iff t \preceq u \wedge \exists i \in
  \mleft[0,\mleft|\stopsOn{t}\mright|\mright):
  \arrTime{t}{\nthStopOn{t}{i}} < \arrTime{u}{\nthStopOn{u}{i}}\,\text{.}
\end{equation}
If two trips have the same stop sequence, but cannot be ordered (because one overtakes the other), we assign them to different lines.
We denote the line of a trip $t$ by $\lineOf{t}$ and define $\stopsOn{\lineOf{t}} = \stopsOn{t}$.
We also define the set of lines at stop $p$ as
\begin{equation}
  \linesAt{p} = \mleft\{ (L,i) \;\middle|\; p = \nthStopOn{L}{i}
  \:\text{where $L$ is a line and}\: \stopsOn{L} =
  \seq{\nthStopOn{L}{0},\nthStopOn{L}{1},\dotsc}\mright\}\,\text{.}
\end{equation}
A \emph{trip segment} $\tripSegment{t}{b}{e}$ represents a trip $t$ traveling from stop $\nthStopOn{t}{b}$ to stop $\nthStopOn{t}{e}$.
A \emph{transfer} between trips $t$ and $u$ ($t \neq u$) is denoted by $\transfer{t}{e}{u}{b}$, where passengers exit $t$ at the $e$th stop and board $u$ at the $b$th.
For all transfers,
\begin{equation}\label{eq:transfer}
  \transfer{t}{e}{u}{b} \implies
    \arrTime{t}{e} + \fpTime{\nthStopOn{t}{e}}{\nthStopOn{u}{b}}
    \leq \depTime{u}{b}
\end{equation}
must hold.
Finally, a \emph{journey} is a sequence of alternating trip segments and transfers, with optional footpaths at the beginning and end.
Each leg of a journey must begin at the stop where the previous one ended.

We consider two well-known problems.
Since both of them are multi-criteria problems, the results are \emph{Pareto sets} representing non-dominated journeys.
A journey dominates another if it is no worse in any criterion; if they are equal in every criterion, we break ties arbitrarily.
Although multi-criteria Pareto optimization is NP-hard in general, it is efficiently tractable for natural criteria in public transit networks~\cite{Muller-Hannemann2006}.
In the \emph{earliest arrival problem}, we are given a source stop $\sourcestop$, a target stop $\targetstop$, and a departure time $\tau$.
The result is a Pareto set of tuples $(\tau_\mathrm{jarr},n)$ of arrival time and number of transfers taken during non-dominated journeys from $\sourcestop$ to $\targetstop$ that leave no earlier than $\tau$.
For the \emph{profile problem}, we are given source stop $\sourcestop$, target stop $\targetstop$, an earliest departure time $\tau_\mathrm{edt}$, and a latest departure time $\tau_\mathrm{ldt}$.
Here, we are asked to compute a Pareto set of tuples $(\tau_\mathrm{jdep},\tau_\mathrm{jarr},n)$ representing non-dominated journeys between $\sourcestop$ and $\targetstop$ with $\tau_\mathrm{edt} \leq \tau_\mathrm{jdep} \leq \tau_\mathrm{ldt}$.
Note that for Pareto-optimality, later departure times are considered to be better than earlier ones.

\subsection{Related work}

Some existing approaches solve these problems by modeling timetable information as a graph, using either the \emph{time-expanded} or the \emph{time-dependent} model.
In the (simple) time-expanded model, a node is introduced for each event, such as a train departing or arriving at a station.
Edges are then added to connect nodes on the same trip, as well as between nodes belonging to the same stop (corresponding to a passenger waiting for the next train).
To model minimum change times, additional nodes and edges are required~\cite{Pyrga2008}.
One advantage of this model is that all edge weights are constant, which allows the use of speedup techniques developed for road networks, such as contraction.
Unfortunately, it turns out that due to different network structures, these techniques do not perform as well for public transit networks~\cite{Berger2009}.
Also, time-expanded graphs are rather large.

The time-dependent approach produces much smaller graphs in comparison.
In the simple model, nodes correspond to stops.
Edges no longer have constant weight, but are instead associated with (piecewise linear) travel time functions, which map departure times to travel times (or, equivalently, arrival times).
The weight then depends on the time at which this function is evaluated.
This model can be extended to allow for minimum change times by adding a node for each line at each stop~\cite{Pyrga2008}.
Some speedup techniques have been applied successfully to time-dependent graphs, such as ALT~\cite{Cionini2014} and Contraction~\cite{Geisberger2010}, although not for multi-criteria problems.
For these, several extensions to Dijkstra's algorithm exist, among them the \emph{Multicriteria Label-Setting}~\cite{Hansen1980}, the \emph{Multi-Label Correcting}~\cite{Dean1999}, the \emph{Layered Dijkstra}~\cite{Brodal2004}, and the \emph{Self-Pruning Connection Setting}~\cite{Delling2010} algorithms.
However, as Dijkstra-variants, each of them has to perform rather costly priority queue operations.

Other approaches do not use graphs at all.
\emph{RAPTOR} (Round-bAsed Public Transit Optimized Router)~\cite{Delling2012} is a dynamic program.
In each round, it computes earliest arrival times for journeys with $n$ transfers, where $n$ is the current round number.
It does this by scanning along lines and, at each stop, checking for the earliest trip of that line that can be reached.
It outperforms Dijkstra-based approaches in practice.
The \emph{Connection Scan Algorithm}~\cite{Dibbelt2013} operates on \emph{elementary connections} (trip segments of length $1$).
It orders them by departure time into a single array.
During queries, this array is then scanned once, which is very fast in practice due to the linear memory access pattern.

A number of speedup techniques have been developed for public transit routing.
\emph{Transfer Patterns}~\cite{Bast2010,Bast2014a} is based on the observation that for many optimal journeys, the sequence of stops where transfers occur is the same.
By precomputing these transfer patterns, journeys can be computed very quickly at query time.
\emph{Public Transit Labeling}~\cite{Delling2015} applies recent advances in hub labeling to public transit networks, resulting in very fast query times.
Another example is the \emph{Accelerated Connection Scan Algorithm}~\cite{Strasser2014}, which combines CSA with multilevel overlay graphs to speed up queries on large networks.
The algorithm presented in this work, however, is a new base algorithm; development of further speedup techniques is a subject for future research.

\section{Algorithm}\label{sec:algorithm}

\subsection{Preprocessing}\label{sec:preprocessing}

We precompute transfers so they can be looked up quickly during queries.
A key observation is that the majority of possible transfers is not needed in order to find Pareto-optimal journeys, and can be safely discarded.
Preprocessing is divided into several steps: Initial computation and reduction.
Initial computation of transfers is relatively straightforward.
For each trip $t$ and each stop $\nthStopOn{t}{i}$ of that trip, we examine $\nthStopOn{t}{i}$ and all stops reachable via (direct) footpaths from $\nthStopOn{t}{i}$.
For each of these stops $q$, we iterate over $\mleft(L,j\mright) \in \linesAt{q}$ and find the first trip $u$ of line $L$ such that a valid transfer $\transfer{t}{i}{u}{j}$ satisfying (\ref{eq:transfer}) exists.
Since, by definition, trips do not overtake other trips of the same line, we can discard any transfers to later trips of line~$L$.
Additionally, we do not add any transfers from the first stop ($i = 0$) or to the last stop ($j = \mleft|\stopsOn{L}\mright| - 1$) of a trip.
Furthermore, transfers to trips of the same line are only kept if either $u \prec t$ or $j < i$; otherwise, it is better to simply remain in the current trip.
See Algorithm~\ref{alg:initial_transfer} for a pseudocode description of this.

\begin{algorithm}[!t]
  \caption{Initial transfer computation}
  \label{alg:initial_transfer}
  \begin{algorithmic}[1]
    \Require Timetable data
    \Ensure Transfer set $T$
    \State $T \gets \varnothing$
    \ForEach{trip $t$}
      \ForEach{stop $\nthStopOn{t}{i}$ on trip $t$ with $i > 0$}
        \ForEach{stop $q$ such that $\fpTime{\nthStopOn{t}{i}}{q}$ is defined}
          \ForEach{line $(L,j) \in \linesAt{q}$ with
              $j < \mleft|\stopsOn{L}\mright| - 1$}
            \State $u \gets$ earliest trip of line $L$ such that $\arrTime{t}{i} +
              \fpTime{\nthStopOn{t}{i}}{q} \leq \depTime{u}{j}$
            \If{$L \neq \lineOf{t} \vee u \prec t \vee j < i$}
              \State $T \gets T \cup \mleft(\transfer{t}{i}{u}{j}\mright)$
              \Comment Add transfer
            \EndIf
          \EndFor
        \EndFor
      \EndFor
    \EndFor
  \end{algorithmic}
\end{algorithm}

\begin{algorithm}[!t]
  \caption{Remove U-turn transfers}
  \label{alg:uturn_transfers}
  \begin{algorithmic}[1]
    \Require Timetable data, transfer set $T$
    \Ensure Reduced transfer set $T$
    \ForEach{transfer $\transfer{t}{i}{u}{j} \in T$}
      \If{$\nthStopOn{t}{i-1} = \nthStopOn{u}{j+1} \wedge
          \arrTime{t}{i-1} + \chTime{\nthStopOn{t}{i-1}}
          \leq \depTime{u}{j+1}$}
        \State $T \gets T \setminus \transfer{t}{i}{u}{j}$
        \Comment Remove U-turn transfer
      \EndIf
    \EndFor
  \end{algorithmic}
\end{algorithm}

After initial computation is complete, we perform a number of reduction steps, where we discard transfers that are not necessary to find Pareto-optimal journeys.
First, we discard any transfers $\transfer{t}{i}{u}{j}$ where $\nthStopOn{u}{j+1} = \nthStopOn{t}{i-1}$ (we call these \emph{U-turn transfers}) as long as
\begin{equation}\label{eq:uturn}
  \arrTime{t}{i-1} + \chTime{\nthStopOn{t}{i-1}} \leq \depTime{u}{j+1}
\end{equation}
holds (Algorithm~\ref{alg:uturn_transfers}).
In this case, we can already reach $u$ from $t$ at the previous stop, and because
\begin{equation}
\begin{alignedat}{3}
  &\arrTime{t}{i-1}\quad&\leq\quad&\depTime{t}{i-1}
    \quad&\leq\quad&\arrTime{t}{i}\\
  \leq\;&\depTime{u}{j}\quad&\leq\quad&\arrTime{u}{j+1}
  \quad&\leq\quad&\depTime{u}{j+1}\,\text{,}
\end{alignedat}
\end{equation}
all trips that can reach $t$ at the previous stop can also reach $u$, and all trips reachable from $u$ are also reachable from $t$.
Equation~(\ref{eq:uturn}) may not hold if the stops in question have different minimum change times.

\begin{algorithm}[tb]
  \caption{Transfer reduction}
  \label{alg:transfer_reduction}
  \begin{algorithmic}[1]
    \Require Timetable data, transfer set $T$
    \Ensure Reduced transfer set $T$
    \ForEach{trip $t$}
      \State $\tau_\mathrm{A}(\cdot) \gets \infty$
      \Comment Arrival time at stops
      \State $\tau_\mathrm{C}(\cdot) \gets \infty$
      \Comment Earliest change time at stops
      \For{$i \gets \mleft|\stopsOn{t}\mright|-1,\dotsc,1$}
        \State $\tau_\mathrm{A}\mleft(\nthStopOn{t}{i}\mright) \gets
          \min\mleft(\tau_\mathrm{A}\mleft(\nthStopOn{t}{i}\mright),
            \arrTime{t}{i}\mright)$
        \ForEach{stop $q$ such that $\fpTime{\nthStopOn{t}{i}}{q}$ is defined}
          \State $\tau_\mathrm{A}(q) \gets \min\mleft(\tau_\mathrm{A}(q),
            \arrTime{t}{i} + \fpTime{\nthStopOn{t}{i}}{q}\mright)$
          \State $\tau_\mathrm{C}(q) \gets \min\mleft(\tau_\mathrm{C}(q),
            \arrTime{t}{i} + \fpTime{\nthStopOn{t}{i}}{q}\mright)$
        \EndFor
        \ForEach{transfer $\transfer{t}{i}{u}{j} \in T$}
          \State $\mathit{keep} \gets \mathbf{false}$
          \ForEach{stop $\nthStopOn{u}{k}$ on trip $u$ with $k > j$}
            \State $\mathit{keep} \gets \mathit{keep} \vee \arrTime{u}{k} <
              \tau_\mathrm{A}\mleft(\nthStopOn{u}{k}\mright)$
            \State $\tau_\mathrm{A}\mleft(\nthStopOn{u}{k}\mright) \gets
              \min\mleft(\tau_\mathrm{A}\mleft(\nthStopOn{u}{k}\mright),
                \arrTime{u}{k}\mright)$
            \ForEach{stop $q$ such that $\fpTime{\nthStopOn{u}{k}}{q}$ is defined}
              \State $\eta \gets \arrTime{u}{k} + \fpTime{\nthStopOn{u}{k}}{q}$
              \State $\mathit{keep} \gets \mathit{keep} \vee \eta <
              \tau_\mathrm{A}(q) \vee \eta < \tau_\mathrm{C}(q)$
              \State $\tau_\mathrm{A}(q) \gets \min\mleft(\tau_\mathrm{A}(q),
                \eta\mright)$
              \State $\tau_\mathrm{C}(q) \gets \min\mleft(\tau_\mathrm{C}(q),
                \eta\mright)$
            \EndFor
          \EndFor
          \If{$\neg\mathit{keep}$}
            \State $T \gets T \setminus \transfer{t}{i}{u}{j}$
            \Comment No improvement, remove transfer
          \EndIf
        \EndFor
      \EndFor
    \EndFor
  \end{algorithmic}
\end{algorithm}

Next, we further reduce the number of transfers by analyzing which transfers lead to improved arrival times.
We do this by moving backwards along a trip, keeping track of where and when passengers in that trip can arrive, either by simply exiting the trip or by transferring to another trip reachable from their current position.
Again, we iterate over all trips $t$.
For each trip, we maintain two mappings $\tau_\mathrm{A}$ and $\tau_\mathrm{C}$ from stops to arrival time and earliest change time, respectively.
Initially, they are set to $\infty$ for all stops.
During execution of the algorithm, they are updated to reflect when passengers arrive ($\tau_\mathrm{A}$) or can board the next trip ($\tau_\mathrm{C}$) at each stop.%
\footnote{If there are no minimum change times, then $\tau_\mathrm{A} = \tau_\mathrm{C}$ and we only maintain $\tau_\mathrm{A}$.}
We then iterate over stops $\nthStopOn{t}{i}$ of trip $t$ in decreasing index order, meaning we examine later stops first.
At each stop, we update the arrival time and change time for that stop if they are improved:
\begin{align*}
\tau_\mathrm{A}\mleft(\nthStopOn{t}{i}\mright) &\gets
  \min\mleft(\tau_\mathrm{A}\mleft(\nthStopOn{t}{i}\mright),
  \arrTime{t}{i}\mright)\quad\text{and}\\
\tau_\mathrm{C}\mleft(\nthStopOn{t}{i}\mright) &\gets
  \min\mleft(\tau_\mathrm{C}\mleft(\nthStopOn{t}{i}\mright),
  \arrTime{t}{i} + \chTime{\nthStopOn{t}{i}}\mright)\,\text{.}
\end{align*}
Similarly, we update $\tau_\mathrm{A}$ and $\tau_\mathrm{C}$ for all stops $q$ reachable via footpaths from $\nthStopOn{t}{i}$:
\begin{align*}
\tau_\mathrm{A}(q) &\gets
  \min\mleft(\tau_\mathrm{A}(q), \arrTime{t}{i} +
  \fpTime{\nthStopOn{t}{i}}{q}\mright)\quad\text{and}\\
\tau_\mathrm{C}(q) &\gets
  \min\mleft(\tau_\mathrm{C}(q), \arrTime{t}{i} +
  \fpTime{\nthStopOn{t}{i}}{q}\mright)\,\text{.}
\end{align*}
We then determine, for each transfer $\transfer{t}{i}{u}{j}$ from $t$ at that stop, if $u$ improves arrival and/or change times for any stop.
To do this, we iterate over all stops $\nthStopOn{u}{k}$ of $u$ with $k > j$ and perform the same updates to $\tau_\mathrm{A}$ and $\tau_\mathrm{C}$ as we did above, this time for $\nthStopOn{u}{k}$ and all stops reachable via footpaths from $\nthStopOn{u}{k}$.
If this results in any improvements to either $\tau_\mathrm{A}$ or $\tau_\mathrm{C}$, we keep the transfer, otherwise we discard it.
Discarded transfers are not required for Pareto-optimal journeys, since we have shown that (a) taking later transfers (or simply remaining in the current trip) leads to equal or better arrival times ($\tau_\mathrm{A}$), and (b) all trips reachable via that transfer can also be reached via those later transfers ($\tau_\mathrm{C}$).
Refer to Algorithm~\ref{alg:transfer_reduction} for a pseudocode description.

All these algorithms are trivially parallelized, since each trip is processed independently.
Also, there is no need to perform them as separate steps; they can easily be merged into one.
We decided to keep them distinct to showcase the separation of concerns.
Furthermore, more complex reduction steps are possible, where there are dependencies between trips.
For example, to minimize the size of the transfer set, one could compute full profiles between all stops (all-to-all), then keep only those transfers required for optimal journeys.
However, that would be computationally expensive.
In contrast, the comparatively simple computations presented here can be performed within minutes, even for large networks, while still resulting in a greatly reduced transfer set (see Section~\ref{sec:experiments} for details).

Note that this explicit representation of transfers allows fine-grained control over them.
For instance, one can easily introduce transfers between specific trips that would otherwise violate the minimum change time or footpath restrictions, or remove transfers from certain trips.
Transfer preferences are another example.
If two trips travel in parallel (for part of their stop sequence), there may be multiple possible transfers between them.
The algorithm described above discards all but the last of them; by modifying it, preference could be given to transfers that are more accessible, for instance.
Since this only has to be considered during preprocessing, query times are unaffected.

\subsection{Earliest Arrival Query}\label{sec:earliest_arrival}

As a reminder, the input to an earliest arrival query consists of the source stop $\sourcestop$, the target stop $\targetstop$, and the (earliest) departure time $\tau$, and the objective is to calculate a Pareto set of $\mleft(\tau_\mathrm{jarr},n\mright)$ tuples representing Pareto-optimal journeys arriving at time $\tau_\mathrm{jarr}$ after $n$ transfers.
During the algorithm, we remember which parts of each trip $t$ have already been processed by maintaining the index $\reached{t}$ of the first reached stop, initialized to $\reached{t} \gets \infty$ for all trips.
We also use a number of queues $Q_n$ of trip segments reached after $n$ transfers and a set $\mathcal{L}$ of tuples $\mleft(L,i,\Delta\tau\mright)$.
The latter indicates lines reaching the target stop $\targetstop$, and is computed by
\begin{multline*}
  \mathcal{L} = \mleft\{\mleft(L,i,0\mright)\,\middle|\,\mleft(L,i\mright) \in
    \linesAt{\targetstop}\mright\}\\
    \cup \mleft\{\mleft(L,i,\fpTime{q}{\targetstop}\mright)\,\middle|\,
    \mleft(L,i\mright) \in \linesAt{q} \wedge \exists\text{ a footpath from
    $q$ to $\targetstop$}\mright\}
\end{multline*}
We start by identifying the trips travelers can reach from $\sourcestop$ at time $\tau$.
For this, we examine $\sourcestop$ and all stops reachable via footpaths from $\sourcestop$.
For each of these stops $q$, we iterate over $\mleft(L,i\mright) \in \linesAt{q}$ and find the first trip $t$ of line $L$ such that
\begin{equation*}
  \depTime{t}{i} \geq \begin{cases}
    \tau &\text{if $q = \sourcestop$,}\\
    \tau + \fpTime{\sourcestop}{q} &\text{otherwise.}
  \end{cases}
\end{equation*}
For each of those trips, if $i < \reached{t}$, we add the trip segment $\tripSegment{t}{i}{\smash{\reached{t}}}$ to queue $Q_0$ and then update $\reached{u} \gets \min\mleft(\reached{u},i\mright)$ where $t \preceq u \wedge \lineOf{t} = \lineOf{u}$, meaning we update the first reached stop for $t$ and all later trips of the same line.
Due to the way $\preceq$ is defined in (\ref{eq:trip_order}), none of these later trips $u$ can improve upon $t$.
By marking them as reached, we eliminate them from the search and avoid redundant work.

\begin{algorithm}[!t]
  \caption{Earliest arrival query}
  \label{alg:ea_query}
  \begin{algorithmic}[1]
    \Require Timetable, transfer set $T$, source stop $\sourcestop$, target stop
    $\targetstop$, departure time $\tau$
    \Ensure Result set $J$
    \State $J \gets \varnothing$
    \State $\mathcal{L} \gets \varnothing$
    \State $Q_n \gets \varnothing$ for $n = 0,1,\dotsc$
    \State $\reached{t} \gets \infty$ for all trips $t$
    \ForEach{stop $q$ such that $\fpTime{q}{\targetstop}$ is defined}
      \State $\Delta\tau \gets 0$ if $\targetstop = q$,
        else $\fpTime{q}{\targetstop}$
      \ForEach{$(L, i) \in \linesAt{q}$}
        \State $\mathcal{L} \gets \mathcal{L} \cup
          \mleft\{\mleft(L, i, \Delta\tau\mright)\mright\}$
      \EndFor
    \EndFor
    \ForEach{stop $q$ such that $\fpTime{\sourcestop}{q}$ is defined}
      \State $\Delta\tau \gets 0$ if $\sourcestop = q$,
        else $\fpTime{\sourcestop}{q}$
      \ForEach{$(L, i) \in \linesAt{q}$}
        \State $t \gets$ earliest trip of $L$ such that $\tau + \Delta\tau \leq
          \depTime{t}{i}$
        \State \textsc{enqueue}$(t, i, 0)$
      \EndFor
    \EndFor
    \State $\tau_\mathrm{min} \gets \infty$
    \State $n \gets 0$
    \While{$Q_n \neq \varnothing$}
      \ForEach{$\tripSegment{t}{b}{e} \in Q_n$}
        \ForEach{$\mleft(\lineOf{t},i,\Delta\tau\mright) \in \mathcal{L}$ with
            $b < i$ and $\arrTime{t}{i} + \Delta\tau < \tau_\mathrm{min}$}
          \State $\tau_\mathrm{min} \gets \arrTime{t}{i} + \Delta\tau$
          \State $J \gets J \cup \mleft\{\mleft(\tau_\mathrm{min}, n
            \mright)\mright\}$, removing dominated entries
        \EndFor
        \If{$\arrTime{t}{b + 1} < \tau_\mathrm{min}$}
          \ForEach{transfer $\transfer{t}{i}{u}{j} \in T$ with $b < i \leq e$}
            \State \textsc{enqueue}$(u, j, n + 1)$
          \EndFor
        \EndIf
      \EndFor
      \State $n \gets n + 1$
    \EndWhile
  \end{algorithmic}
  \hfill
  \begin{algorithmic}[1]
    \Procedure{enqueue}{trip $t$, index $i$, number of transfers $n$}
      \If{$i < \reached{t}$}
        \State $Q_n \gets Q_n \cup
          \mleft\{\tripSegment{t}{i}{\reached{t}}\mright\}$
        \ForEach{trip $u$ with $t \preceq u \wedge \lineOf{t} = \lineOf{u}$}
          \State $\reached{u} \gets \min\mleft(\reached{u},i\mright)$
        \EndFor
      \EndIf
    \EndProcedure
  \end{algorithmic}
\end{algorithm}

After the initial trips have been found, we operate on the trip segments in $Q_0, Q_1, \dotsc$ until there are no more unprocessed elements.
For each trip segment $\tripSegment{t}{b}{e} \in Q_n$, we perform the following three steps.
First, we check if this trip reaches the target stop.
For each $\mleft(\lineOf{t},i,\Delta\tau\mright) \in \mathcal{L}$ with $i > b$, we generate a tuple $\mleft(\arrTime{t}{i} + \Delta\tau, n\mright)$ and add it to the result set, maintaining the Pareto property.
Second, we check if this trip should be pruned because it cannot lead to a non-dominated journey.
This is the case if we already found a journey with $\tau_\mathrm{jarr} < \arrTime{t}{b + 1}$.
Third, if the trip is not pruned, we examine its transfers.
For each transfer $\transfer{t}{i}{u}{j}$ with $b < i \leq e$, we check if $j < \reached{u}$.
If so, we add $\tripSegment{u}{j}{\smash{\reached{u}}}$ to $Q_{n+1}$ and update $\reached{v} \gets \min\mleft(\reached{v},j\mright)$ for all $v$ with $u \preceq v \wedge \lineOf{u} = \lineOf{v}$.
Otherwise, we already reached $u$ or an earlier trip of the same line at $j$ or an earlier stop, and we skip the transfer.
A pseudocode description can be found in Algorithm~\ref{alg:ea_query}.

The main loop is similar to a breadth-first search: First, all trips reachable directly from the source stop are examined, then all trips reached after a transfer from those, etc.
Therefore, we find journeys with the least number of transfers first.
Any non-dominated journey discovered later cannot have a lower number of transfers and must therefore arrive earlier.
This property enables the pruning in step two, which prevents us from having to examine all reachable trips regardless of the target.
However, it also means that the journey with the earliest arrival time is the last one discovered, and all journeys with less transfers are found beforehand.
This is why we only consider the multi-criteria problem variants.

\subsection{Profile Query}

We perform profile queries by running the main loop of an earliest arrival query for each distinct departure time in the given interval, preserving labels between runs to avoid redundant work.
Later journeys dominate earlier journeys, provided arrival time and number of transfers are equal or better, while earlier journeys never dominate later ones.
Therefore, we process later departures first.
However, in order to reuse labels across multiple runs, we need to keep multiple labels for each trip, consisting of the index of the first reached stop and the number of transfers required to reach it.
Since the number of transfers is limited in practice, we use $\reached[n]{t}$ to denote the first stop reached on trip $t$ after at most $n$ transfers and update $\reached[n+1]{t}$ (and following) whenever we update $\reached[n]{t}$.
To decide if a trip segment should be queued while processing $Q_n$, we compare against and update $\reached[n+1]{t}$.
We also change the pruning step so we compare against the minimum arrival time of journeys with no more than $n + 1$ transfers.

To see why labels can be reused, consider two runs with departure times $\tau_1$ and $\tau_2$, where $\tau_1 < \tau_2$, which both reach trip $t$ at stop $i$ after $n$ transfers.
Continuing from this point, both will reach the destination at the same time and after the same number of transfers.
However, since $\tau_1 < \tau_2$, the journeys departing at $\tau_2$ dominate the journeys departing at $\tau_1$.
Knowing this, we can avoid computing them in the first place by computing $\tau_2$ first and keeping the labels.

\subsection{Implementation}\label{sec:implementation}

We improve the performance of the algorithm by taking advantage of SIMD (single instruction, multiple data) instructions, avoiding dynamic memory allocations and increasing locality of reference (reducing cache misses).
In our data instances, all lines have less than $200$ stops.
Also, none of our tests found Pareto-optimal journeys with $16$ or more transfers.
Thus, we set the maximum number of transfers to $15$.
During profile queries, we can then update $\reached[0]{t}$ to $\reached[15]{t}$ using a single $128$-bit vector minimum operation.

To avoid memory allocations during query execution, we replace the $n$ queues with a single, preallocated array.
To see why this is possible, note that the maximum number of trip segments queued is bounded by the number of elementary connections.
We use pointers to keep track of the current element, the end of the queue, and the level boundaries (where the number of transfers $n$ is increased).

We improve locality of reference by splitting the steps of the inner loop into three separate loops.
Thus, we iterate three times over each level, each time updating the elements in the ``queue'', before increasing $n$ and moving on to the next level.
In the first iteration, we look up $\arrTime{t}{b+1}$ and store it next to the trip segment into the queue.
Additionally, we check $\mathcal{L}$ to see if the trip reaches the destination, and update arrival times as necessary.
In the second iteration, we perform the pruning step by comparing the time stored in the queue with the arrival time at the destination.
If the element is not pruned, we replace it with two indices into the array of transfers, indicating the transfers corresponding to the trip segment.
If the element is pruned, we set both indices to $0$, resulting in an empty interval.
Finally, in the third iteration, we examine this list of transfers and add new trip segments to the queue as necessary.
Thus, arrival times $\arrTime{\cdot}{\cdot}$ are required only in the first loop, transfer indices only in the second loop, and transfers and reached stops $\reached[n]{t}$ only in the final loop.
This leads to reduced cache pressure and therefore to less cache misses, which in turn results in improved performance (see Section~\ref{sec:experiments}).
For more details on the data structures used, please refer to Appendix~\ref{app:data_structures}.

\subsection{Journey Descriptions}\label{sec:journey_descriptions}

So far, we only described how to compute arrival time and number of transfers of journeys, which is enough for many applications.
However, we can retrieve the full sequence of trip segments as follows.
Whenever a trip segment is queued, we store with it a pointer to the currently processed trip segment.
Since we replaced the queue with a preallocated array, all entries are preserved until the end of the query.
Therefore, when we find a journey reaching the destination, we simply follow this chain of pointers to reconstruct the sequence of trip segments.
If required, the appropriate transfers between the trips can be found by rescanning the list of transfers.

\section{Experiments}\label{sec:experiments}

We ran experiments on a dual 8-core Intel~Xeon~E5-2650\,v2 processor clocked at 2.6\,GHz, with 128\,GB of DDR3-1600 RAM and 20\,MB of L3~cache.
Our code was compiled using g++~4.9.2 with optimizations enabled.
We used two test instances, summarized in Table~\ref{tab:instances}.
The first, available at \texttt{data.london.gov.uk}, covers Greater London and includes data for underground, bus, and Docklands Light Railway services for one day.
The second consists of data used by \texttt{bahn.de} during winter 2011/2012, containing European long distance trains, German local trains, and many buses over two days.

Table~\ref{tab:instances} also reports the number of transfers before and after reduction, as well as the total space consumption (for the reduced transfers and all timetable data).
Reduction eliminates about $84\%$ of transfers for London, and almost $90\%$ for Germany.
The times required for preprocessing can be found in Table~\ref{tab:preprocessing}.

\begin{table}[t]
  \centering
  \caption{Instances used for experiments}\label{tab:instances}
  \pgfplotstabletypeset[
    columns={[index]0,london,db},
    display columns/0/.style={column name={}, string type, column type=l,
    string replace={Mem. Consumption}{\midrule{}Space consumption},
    },
    columns/london/.style={column name=London, int detect, column type=r,
      assign cell content/.code={
        \ifnum\pgfplotstablerow=7%
          \bytesToMiB{##1}%
        \else
          \pgfkeyssetvalue{/pgfplots/table/@cell content}{%
            \pgfkeysalso{/pgf/number format/int detect}%
            \pgfmathprintnumber{##1}%
          }
        \fi
      },
    },
    columns/db/.style={column name=Germany, int detect, column type=r,
      assign cell content/.code={
        \ifnum\pgfplotstablerow=7%
          \bytesToMiB{##1}%
        \else
          \pgfkeyssetvalue{/pgfplots/table/@cell content}{%
            \pgfkeysalso{/pgf/number format/int detect}%
            \pgfmathprintnumber{##1}%
          }
        \fi
      },
    },
  ]{data/instances_manual.table}
\end{table}

\begin{table}[t]
  \centering
  \caption{Preprocessing times for transfer computation and reduction}%
  \label{tab:preprocessing}
  \pgfplotstabletypeset[
    columns={[index]0,london-1,london-16,db-1,db-16},
    display columns/0/.style={
      column name={}, string type, column type=l,
      string replace={Total}{\midrule{}Total}
    },
    columns/london-1/.style={
      column name=\mlcell{London\\$1$ thread},
      int detect, column type=r,
      postproc cell content/.append style={
        /pgfplots/table/@cell content/.add={}{\,s}
      }},
    columns/london-16/.style={
      column name=\mlcell{London\\$16$ threads},
      int detect, column type=r,
      postproc cell content/.append style={
        /pgfplots/table/@cell content/.add={}{\,s}
      }},
    columns/db-1/.style={
      column name=\mlcell{Germany\\$1$ thread},
      int detect, column type=r,
      postproc cell content/.append style={
        /pgfplots/table/@cell content/.add={}{\,s}
      }},
    columns/db-16/.style={
      column name=\mlcell{Germany\\$16$ threads},
      int detect, column type=r,
      postproc cell content/.append style={
        /pgfplots/table/@cell content/.add={}{\,s}
      }},
  ]{data/preprocessing.table}
\end{table}

Running times reported for queries are averages over $10\,000$ queries with source and target stops selected uniformly at random.
For profile queries, the departure time range is the first day covered by the timetable; for earliest arrival queries, the departure time is selected uniformly at random from that range.
We do not compute full journey descriptions.

We evaluated the optimizations described in Section~\ref{sec:implementation}, as well as the effect of transfer reduction, on the London instance (Table~\ref{tab:variants}).
SIMD instructions are only used in profile queries and enabling them has no effect on earliest arrival queries.
With all optimizations, running time for profile queries is improved by a factor of $2$.
Transfer reduction improves running times by a factor of $3$.

\begin{table}[p]
  \centering
  \caption{Evaluation of optimizations in Section~\ref{sec:implementation},
    using the London instance}\label{tab:variants}
  \pgfplotstabletypeset[
    columns={[index]0,earliest,profile},
    display columns/0/.style={
      column name=variant, string type, column type=l,
      string replace={nosse}{Basic, without SIMD},
      string replace={singleloop}{Basic, with SIMD},
      string replace={moreasm}{Optimized},
      string replace={full-moreasm}{\midrule{}Optimized, all transfers},
    },
    columns/earliest/.style={
      column name=earliest arr.\ (ms), column type=r,
      divide by=1000, fixed, precision=1, fixed zerofill,
    },
    columns/profile/.style={
      column name=profile (ms), column type=r, int detect,
      divide by=1000, fixed, precision=1, fixed zerofill,
    },
  ]{data/variants.table}
\end{table}

\begin{table}[p]
  \centering
  \caption{Comparison with the state of the art. Results taken from
    \protect{\cite{Bast2014,Bast2014a,Delling2015,Strasser2014}}. Bicriteria
    algorithms computing a set of Pareto-optimal journeys regarding arrival
    time and number of transfers are marked in column ``tr.'' (others only
    optimize arrival time). Profile queries are marked in column ``pr.''.
    }\label{tab:comparison}
  \pgfplotstabletypeset[
    columns={Algorithm,Instance,Stops,Connections,Transfers,Profile,
      Preprocessing,Comparisons,Time},
    display columns/0/.style={string type, column type=l,
      column name=algorithm,
      string replace={ours}{TripBased},
    },
    columns/Instance/.style={string type,column type=l,
      column name=instance,
      string replace={london}{London},
      string replace={db}{Germany},
    },
    columns/Stops/.style={column type=r,
      column name=\kern-2em\mlcell{stops\\($\cdot 10^3$)},
      fixed, precision=1, fixed zerofill, divide by=1000,
    },
    columns/Connections/.style={column type=r,
      column name=\kern-1em\mlcell{conn.\\($\cdot 10^6$)},
      fixed, precision=1, fixed zerofill, divide by=1000000,
    },
    columns/Transfers/.style={string type,column type=c,
      column name=tr.,
      string replace={0}{$\circ$},
      string replace={1}{$\bullet$},
    },
    columns/Profile/.style={string type,column type=c,
      column name=\kern-1em{}pr.,
      string replace={0}{\kern-1em$\circ$},
      string replace={1}{\kern-1em$\bullet$},
    },
    columns/Preprocessing/.style={column type=r,
      column name=\mlcell{prep.\\(h)}, string type,
      string replace={0.1}{$<0.1$},
      string replace={0}{}, empty cells with={---},
    },
    columns/Comparisons/.style={
      column name=\mlcell{comp.\\/stop}, column type=r,
      string replace={0}{}, empty cells with={n/a},
      fixed, precision=1, fixed zerofill,
    },
    columns/Time/.style={
      column name=\mlcell{query\\(ms)}, column type=r,
      divide by=1000, fixed, precision=1, fixed zerofill,
      dec sep align,
    },
    every row 2 column 8/.style={/pgf/number format/precision=2},
    every row no 6/.style={before row={\midrule}},
    every row no 10/.style={before row={\midrule[\heavyrulewidth]}},
    every row no 14/.style={before row={\midrule}},
  ]{data/query_times_manual.table}
\end{table}

We compare our new algorithm to the state of the art in Table~\ref{tab:comparison}.
We distinguish between algorithms which optimize arrival time only ($\circ$) and those that compute Pareto sets optimizing arrival time and number of transfers ($\bullet$), and between earliest arrival ($\circ$) and profile ($\bullet$) queries.
We report the average number of label comparisons per stop%
\footnote{Note that in our algorithm, labels are not associated with stops, but with trips instead.
  For better comparison with previously published work, we divided the total number of label comparisons by the number of stops.
}, where available, and the average running time.
Direct comparison with the Accelerated Connection Scan Algorithm~(ACSA)~\cite{Strasser2014} and Contraction Hierarchies~(CH)~\cite{Geisberger2010} is difficult, since they do not support bicriteria queries.%
\footnote{ACSA uses transfers to break ties between journeys with equal arrival times.}
We have faster query times than CSA~\cite{Dibbelt2013} and RAPTOR~\cite{Delling2012}, at the cost of a few minutes of preprocessing time.
Transfer Patterns~(TP)~\cite{Bast2010,Bast2014a} and Public Transit Labeling~(PTL)~\cite{Delling2015} have faster query times (especially on larger instances), however, their preprocessing times are several orders of magnitude above ours.

\begin{figure}[t]%
\centering%
\begin{minipage}[b]{0.48\textwidth}%
\iftoggle{haspgfplotsstatistics}{%
  \begin{tikzpicture}
    \begin{semilogyaxis}[small,
        xlabel={geo-rank ($2^r$)},
        ylabel={running time (ms)},
        boxplot/draw position=4+\plotnumofactualtype,
        xtick={4,6,...,16},
        enlarge x limits={abs=0.5},
    ]\input{data/georanked-db-earliest.plot}
    \end{semilogyaxis}
  \end{tikzpicture}%
}{\includegraphics{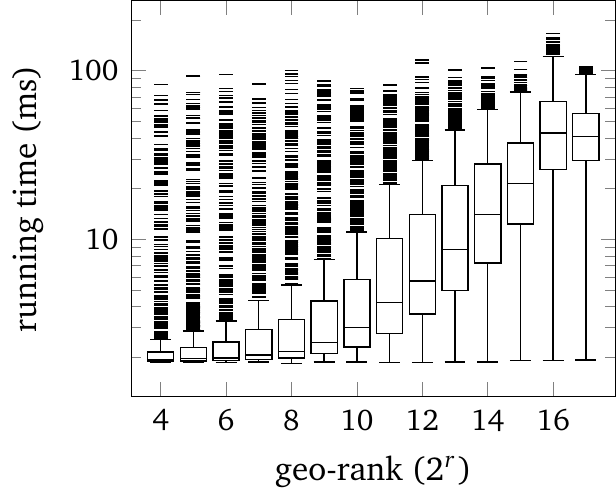}}%
  \caption{Earliest arrival query times by geo-rank on Germany}%
  \label{fig:geo-ea}%
\end{minipage}\hfill%
\begin{minipage}[b]{0.48\textwidth}%
\iftoggle{haspgfplotsstatistics}{%
  \begin{tikzpicture}
    \begin{semilogyaxis}[small,
        xlabel={geo-rank ($2^r$)},
        ylabel={running time (ms)},
        boxplot/draw position=4+\plotnumofactualtype,
        xtick={4,6,...,16},
        enlarge x limits={abs=0.5},
    ]\input{data/georanked-db-profile.plot}
    \end{semilogyaxis}
  \end{tikzpicture}%
}{\includegraphics{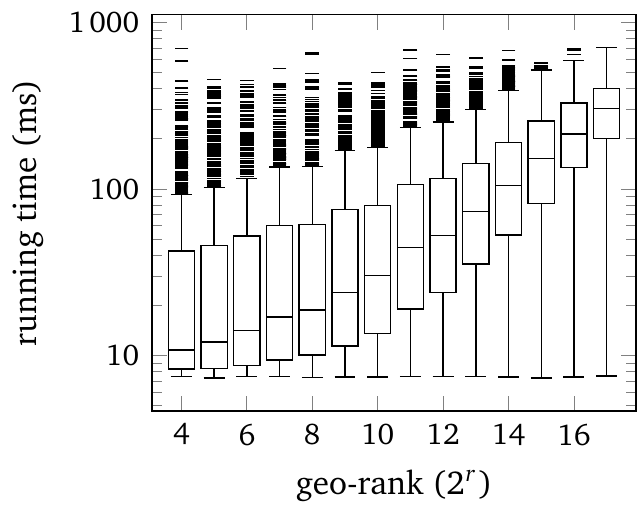}}%
  \caption{Profile query times by geo-rank on Germany}%
  \label{fig:geo-profile}%
\end{minipage}
\end{figure}

To examine query times further, we ran $1\,000$ geo-rank queries~\cite{Strasser2014}.
A geo-rank query picks a stop uniformly at random and orders all other stops by geographical distance.
Queries are run from the source stop to the $2^r$-th stop, where $r$ is the geo-rank.
Results for the Germany instance are reported in Figure~\ref{fig:geo-ea} (earliest arrival queries) and Figure~\ref{fig:geo-profile} (profile queries).
Note the logarithmic scale on both axes.
Query times for the maximum geo-rank are about the same as the average query time when selecting source and target uniformly at random, since randomly selected stops are unlikely to be near each other.
Local queries, which are often more relevant in practice, are generally much faster (by an order of magnitude), although there is a significant number of outliers, since physically close locations do not necessarily have direct or fast connections.

\section{Conclusion}\label{sec:conclusions}

We presented a novel algorithm for route planning in public transit networks.
By focusing on trips and transfers between them, we computed multi-criteria profiles optimizing arrival time and number of transfers on a metropolitan network in $70$\,ms with a preprocessing time of just $30$\,s, occupying a Pareto-optimal spot among current state of the art algorithms.
The explicit representation of transfers allows fine-grained modeling, while the short preprocessing time allows the use in dynamic scenarios.
In addition, localized changes (such as trip delays or cancellations) do not necessitate a full rerun of the preprocessing phase.
Instead, only a subset of the data needs to be updated.
Development of suitable algorithms is a subject of future studies.
Future work also includes efficiently extending the covered period of time by exploiting periodicity in timetables, making the algorithm more scalable by using network decomposition, and extending it to support more generic criteria such as fare zones or walking distance.

\appendix
\bibliographystyle{hplain}
\bibliography{references}

\section{Data Structures}\label{app:data_structures}

We assign consecutive integer IDs, starting from $0$, to stops, lines, and trips.
For trips, we assign IDs such that trips of the same line are consecutive, with earlier trips having lower IDs.
An array maps lines to their first trip.
We store the remaining data using a forward star representation.
For example, to store footpaths, we use two arrays, \texttt{Footpaths} and \texttt{FootpathIndex}.
\texttt{Footpaths} contains information about footpaths, namely the destination stop and the length, for all footpaths.
It is ordered such that footpaths starting at stop $0$ come first, then footpaths starting at stop $1$, etc.
\texttt{FootpathIndex} contains, for each stop, the index of the first footpath starting at that stop (plus a sentinel value equal to the size of \texttt{Footpaths}).
To examine footpaths at stop $i$, we iterate from \texttt{Footpaths[FootpathIndex[$i$]]} to \texttt{Footpaths[FootpathIndex[$i+1$]]}.
The lines at each stop and the stops on each line are similarly stored.
For arrival times, we store all times of the first trip in order, then all times of the second trip etc.\ and store the index of the first entry for each trip.
Thus, $\arrTime{t}{i}$ is implemented by looking up \texttt{ArrivalTimes[TripTimeIndex[$t$]$\vphantom{.}+i$]}.
Departure times use the same index array, as do transfers, albeit with an additional indirection: Transfers from trip $t$ at index $i$ can be found starting at \texttt{Transfers[TransferIndex[TripTimeIndex[$t$]$\vphantom{.}+i$]]}.
For the transfers themselves, we only store the target trip and board index.
Finally, we redundantly store an array mapping trips to lines and a second list of footpaths, this time indexed by their destination stops.
Although not strictly necessary, these allow fast lookups during queries.

\end{document}